\documentclass{jltp}

\usepackage{graphicx} 

\title{
Impurity Effect on Kramer-Pesch Core Shrinkage
in $s$-Wave Vortex and Chiral $p$-Wave Vortex
}

\author{
Nobuhiko Hayashi, Yusuke Kato$^*$, and Manfred Sigrist
}

\address{
Institut f\"ur Theoretische Physik,
ETH-H\"onggerberg,
CH-8093 Z\"urich, Switzerland \\
$^*$Department of Basic Science,
University of Tokyo,
Tokyo 153-8902, Japan
}

\runninghead{N. Hayashi, Y. Kato, and M. Sigrist}
{Impurity Effect on Kramer-Pesch Core Shrinkage}

\begin{document}

\maketitle

\begin{abstract}
   The low-temperature shrinking of the vortex core (Kramer-Pesch effect)
is studied for an isolated single vortex
for chiral $p$-wave and $s$-wave superconducting phases.
   The effect of nonmagnetic impurities on
the vortex core radius
is numerically investigated in the Born limit by means of
a quasiclassical approach.
   It is shown that
in the chiral $p$-wave phase the Kramer-Pesch effect
displays a certain robustness against impurities
owing to a specific quantum effect,
while the $s$-wave phase reacts more sensitively to impurity
scattering.  This suggests chiral $p$-wave superconductors as
promising candidates for the experimental observation of  
the Kramer-Pesch effect.

PACS numbers: 74.25.Op, 74.20.Rp, 74.70.Pq
\end{abstract}

\section{INTRODUCTION}
   Much attention has been focused on
vortices in nature,\cite{vortex1}
especially for the quantized vortex in fermionic superfluid and
superconducting systems.\cite{Rev0,Rev1,Rev2}
   One of the fundamental physical quantities of the quantized vortex
is the radius of vortex core.
   Kramer and Pesch\cite{KP} have pointed out theoretically that
the radius of vortex core
decreases proportionally to the temperature $T$
at low temperatures,
much stronger than anticipated
from the temperature dependence of the coherence length.
   This anomalously strong shrinking of the vortex core,
the so-called Kramer-Pesch (KP) effect,\cite{KP,huebener}
occurs when the fermionic spectrum of
vortex bound states\cite{caroli,hess89,gygi,schopohl2,rainer,haya96,y-tanaka}
crosses the Fermi level.\cite{volovik93}
   The temperature dependence of the vortex core
has been theoretically
investigated in the case of
superconductors.\cite{gygi,PK,brandt,rammer,oka96,golubov97,haya98,m-kato,m-kato2,kato01}
   The low-temperature limit of the vortex core radius
was discussed also for dilute Fermi superfluids\cite{elgaroy01-1}
and superfluid neutron star matter.\cite{elgaroy99}

   There are several length scales
which characterize the radius of
vortex core.\cite{gygi,elgaroy99,sonier04-1}
   One of them is the coherence length $\xi(T)$.
   The pair potential $\Delta(r)$ depressed inside the vortex core
is restored at a distance $r\sim \xi(T)$
away from the vortex center $r=0$.\cite{ovchinnikov}
However, 
$\xi(T)$ is almost temperature independent at low temperatures.
   Another length scale is related to the slope of the pair potential
at the vortex center, which is defined as
\begin{equation}
\frac{1}{\xi_1}
=\frac{1}{\Delta(r \rightarrow \infty)}
\lim_{r \rightarrow 0} \frac{\Delta(r)}{r}.
\label{eq:KP}
\end{equation}
   The KP effect means $\xi_1(T) \propto T $
for $T \to 0$,
while the pair potential $\Delta(r)$ is restored
at a distance $r\sim\xi$ ($\gg\xi_1$) at low temperatures.\cite{ovchinnikov}
   Since the spatial profiles of the pair potential $\Delta(r)$
and the supercurrent density $j(r)$
in the vicinity of the vortex center
are related with each other through the low-energy vortex bound states,
the length $\xi_1$ scales with 
the distance $r=r_0$ at which $|j(r)|$ reaches
its maximum value.\cite{KP,sonier04-1,sonier00}
   Therefore, the KP effect gives rise to 
$r_0 \sim \xi_1 \rightarrow 0$ ($T \rightarrow 0$) linear in $T$.

   Recently, $\mu$SR experiments were performed
on the $s$-wave superconductor NbSe$_2$
to observe the KP effect.\cite{sonier04-1,sonier00,sonier97,miller}
   The experimental data of spin precessions,
which correspond to
the Fourier transformation of
the Redfield pattern
(the magnetic field distribution)
of a vortex lattice,
are fitted by a theoretical formula,\cite{yaouanc}
extracting the information on the spatial profile of
the supercurrent density $j(r)$ around vortices.\cite{sonier04-1,sonier00}
   While a shrinking vortex core was observed, it was weaker than
the theoretical expectation of the KP effect for a clean superconductor.
   That is,
indeed the observed vortex core radius $r_0$
seemingly shrank linearly in $T$,
but was extrapolated towards
a finite value in zero-temperature limit,
indicating a saturation of the KP core shrinkage
at a low temperature.
   The measurement of the vortex core radius $r_0$ by $\mu$SR
was also performed for CeRu$_2$ with a similar result.\cite{kadono01}
   Since the KP effect is directly connected with
the low-energy vortex bound states,
the energy level broadening due to impurities may give rise to
a modification of the KP effect,\cite{KP}
in particular, a saturation as observed.
   Impurities exist inevitably in any solid state material,
so that such
a saturation effect is not unlikely to occur.
   Even it turns out that a rather
small concentration of impurities,
leaving the material moderately clean,
would have a strong influence
as observed
in experiments.\cite{sonier04-1,sonier00,sonier97,miller,kadono01}

   There are several factors which influence
the behavior of vortex core radius.
   (i) Impurity effects.\cite{KP}
   (ii) The discreteness of the energy levels of
the vortex bound states.\cite{KP,haya98,m-kato,m-kato2}
   (iii) Vortex lattice
effects.\cite{sonier04-1,sonier00,golubov94,oka99,miranovic,kogan,sonier97-2}
   (iv) Fermi liquid effects $F_1^{\rm s}$.\cite{fogelstrom}
   (v) Antiferromagnetic correlations induced inside
the vortex core as suggested for
cuprate superconductors.\cite{kadono04}
   (vi) The presence of multiple gaps
in multi-band superconductors.\cite{koshelev,haya}

   In this paper,
   we investigate
the temperature dependence of
the vortex core radius $\xi_1(T)$ incorporating
the effect of nonmagnetic impurities on the level of the Born approximation,
for both a chiral $p$-wave 
and an $s$-wave superconductor.
   For this purpose,
we set up a model
of single vortex in a two-dimensional superconductor, where the chiral 
$p$-wave pairing has the form\cite{rice,maeno}
${\bf d}({\bar {\bf k}})={\bar {\bf z}} ({\bar k}_x \pm i {\bar k}_y)$
as, for example, realized in
Sr$_2$RuO$_4$. 

   The examination of the temperature dependence of $ \xi_1(T) $ suggests
that under certain conditions the chiral $p$-wave state shows a more
robust KP effect against impurities
than an $s$-wave state.
   This behavior is connected with the compensation of the intrinsic
phase structure and the vortex phase winding, if chirality and phase winding
are oppositely
oriented.\cite{kato01,volovik99,kato00,kato02,haya02-1,haya03-1,kato03,matsumoto00,matsumoto01,Heeb99}
   Therefore, we argue that chiral $p$-wave superconductors might be
better candidates for the experimental observation of the KP effect.
The chiral $p$-wave superconductivity has been proposed for
Sr$_2$RuO$_4$
with rather strong experimental evidence,\cite{rice,maeno}
and more
recently for Na$_x$CoO$_{2}\cdot y$H$_{2}$O.\cite{tanaka,han}

   The paper is organized as follows.
   In Sec.\ 2, the self-consistent system of equations
for the superconducting order parameter and the impurity self energy
is formulated
on the basis of the quasiclassical theory of superconductivity.
   In Sec.\ 3, the systems of
the $s$-wave vortex and the chiral $p$-wave vortex are described.
   The numerical results for the vortex core radius $\xi_1(T)$
are shown in Sec.\ 4.
   The summary is given in Sec.\ 5.

%
\section{QUASICLASSICAL THEORY}
   Our theoretical analysis of the KP effect will be based on
the quasiclassical theory of
superconductivity,\cite{Eilenberger,kusunose}
which allows us to take inhomogeneous structures as a vortex into account
and at the same time to deal with impurity scattering
on a straightforward way.

   We start with
the quasiclassical Green function
in the presence of nonmagnetic impurities,
\begin{equation}
{\hat g}(i\omega_n,{\bf r},{\bar{\bf k}})=
-i\pi
\pmatrix{
g &
if \cr
-if^{\dagger} &
-g \cr
},
\label{eq:qcg}
\end{equation}
which is the solution of the Eilenberger equation
\begin{equation}
i {\bf v}_{\rm F}({\bar{\bf k}}) \cdot
{\bf \nabla}{\hat g}
+ \bigl[ i\omega_n {\hat \tau}_{3}-{\hat \Delta}-{\hat \sigma},
{\hat g} \bigr]
=0.
\label{eq:eilen}
\end{equation}
Here, ${\hat \Delta}$ is
the superconducting order parameter
\begin{equation}
{\hat \Delta}({\bf r},{\bar{\bf k}})=
\pmatrix{
0 &
\Delta \cr
-\Delta^{*} &
0 \cr
},
\label{eq:SC}
\end{equation}
and ${\hat \sigma}$ denotes the self energy correction
due to impurity scattering
\begin{equation}
{\hat \sigma}(i\omega_n,{\bf r},{\bar{\bf k}})=
\pmatrix{
\sigma_{11} &
\sigma_{12} \cr
\sigma_{21} &
\sigma_{22} \cr
}.
\label{eq:IMP}
\end{equation}
   The Eilenberger equation
is supplemented by the normalization condition
${\hat g}(i\omega_n,{\bf r},{\bar{\bf k}})^2
=-\pi^2{\hat 1}$.\cite{Eilenberger,kusunose}
      The vector
${\bf r}$ is the real-space coordinates
and
the unit vector
${\bar{\bf k}}$
represents
the sense of
the wave number
on the Fermi surface.
   ${\bf v}_{\rm F}({\bar{\bf k}})$ is the Fermi velocity,
$\omega_n=\pi T (2n+1)$ is the fermionic Matsubara frequency
(with the temperature $T$ and the integer $n$),
${\hat \tau}_{3}$ is the third Pauli matrix
in the $2\times2$ particle-hole space,
and
the commutator $[{\hat a},{\hat b}]={\hat a}{\hat b}-{\hat b}{\hat a}$.
   We will consider
an isolated single vortex
in extreme type-II
superconductors (Ginzburg-Landau parameter $\kappa \gg 1$),
and therefore
the vector potential
is neglected in the Eilenberger equation (\ref{eq:eilen}).
   Throughout the paper, vectors with the upper bar denote unit vectors
and we use units in which $\hbar = k_{\rm B} = 1$.

   We now define an alternative impurity self energy ${\hat \Sigma}$ as
\begin{equation}
{\hat \Sigma}(i\omega_n,{\bf r},{\bar{\bf k}})=
\pmatrix{
\Sigma_{\rm d} &
\Sigma_{12} \cr
\Sigma_{21} &
-\Sigma_{\rm d} \cr
}
=
\pmatrix{
\frac{1}{2}(\sigma_{11}-\sigma_{22}) &
\sigma_{12} \cr
\sigma_{21} &
-\frac{1}{2}(\sigma_{11}-\sigma_{22}) \cr
}.
\label{eq:IMP2}
\end{equation}
   The original impurity self energy (\ref{eq:IMP}) can be expressed as
\begin{equation}
{\hat \sigma}={\hat \Sigma}
+\frac{\sigma_{11}}{2} {\hat 1}
+\frac{\sigma_{22}}{2} {\hat 1}.
\label{eq:IMP3}
\end{equation}
   Hence,
we rewrite the Eilenberger equation (\ref{eq:eilen}) as
\begin{equation}
i {\bf v}_{\rm F}({\bar{\bf k}}) \cdot
{\bf \nabla}{\hat g}
+ \bigl[ i{\tilde \omega}_n {\hat \tau}_{3}-\hat{\tilde \Delta},
{\hat g} \bigr]
=0,
\label{eq:eilen2}
\end{equation}
with  the renormalized Matsubara frequency
\begin{equation}
i{\tilde \omega}_n = i\omega_n - \Sigma_{\rm d},
\label{eq:self-w}
\end{equation}
and the renormalized superconducting order parameter
\begin{equation}
\hat{\tilde \Delta}=
\pmatrix{
0 &
\Delta + \Sigma_{12} \cr
- (\Delta^{*} - \Sigma_{21}) &
0 \cr
}.
\label{eq:self-delta}
\end{equation}

   We restrict here to  $s$-wave scattering at the impurities.
   The single-impurity $t$-matrix\cite{Thuneberg84}
is then calculated as
\begin{eqnarray}
{\hat t}(i\omega_n, {\bf r}) =
v{\hat 1} + N_{0}v
\Bigl\langle {\hat g}(i\omega_n, {\bf r},{\bar {\bf k}}) \Bigr\rangle
{\hat t}(i\omega_n, {\bf r}),
\label{eq:t-matrix}
\end{eqnarray}
where
$v$ is the impurity potential for the $s$-wave scattering channel,
$N_{0}$ is the normal-state density of states at the Fermi level, and
the brackets $\langle \cdots \rangle$ denote
the average over the Fermi surface.
   The impurity self energy ${\hat \sigma}$
is given by
\begin{eqnarray}
{\hat \sigma}(i\omega_n, {\bf r})
=
n_{\rm i} {\hat t}(i\omega_n, {\bf r})
=
\frac{n_{\rm i} v}{D} \Bigl[
{\hat 1} + N_{0} v
\Bigl\langle {\hat g}(i\omega_n, {\bf r},{\bar {\bf k}}) \Bigr\rangle
\Bigr],
\label{eq:imp-self}
\end{eqnarray}
where
the denominator is
\begin{eqnarray}
D=1+(\pi N_0 v)^2 \bigl[
\langle g \rangle^2
+ \langle f \rangle
\langle f^{\dagger} \rangle
\bigr],
\label{eq:denomi}
\end{eqnarray}
and $n_{\rm i}$ is the density of impurities.
   The scattering phase shift $\delta_0$ is defined by
$\tan \delta_0 = -\pi N_0 v$.
   In this paper, we investigate the Born limit
($\delta_0 \ll 1$).
   The impurity self energy (\ref{eq:imp-self}) in this limit becomes
\begin{eqnarray}
{\hat \sigma}(i\omega_n, {\bf r})
   &=&                                                 \nonumber
n_{\rm i} v {\hat 1} + \frac{\Gamma_{\rm n}}{\pi}
\Bigl\langle {\hat g}(i\omega_n, {\bf r},{\bar {\bf k}}) \Bigr\rangle \\
   &=&
n_{\rm i} v {\hat 1} + \Gamma_{\rm n}
\pmatrix{
-i \langle g \rangle &
   \langle f \rangle \cr
 - \langle f^{\dagger} \rangle &
 i \langle g \rangle \cr
},
\label{eq:imp-self2}
\end{eqnarray}
where we have defined the impurity scattering rate in the normal state as
$\Gamma_{\rm n}=1/2\tau_{\rm n}=\pi n_{\rm i} N_0 v^2$.
   The mean free path $l$ is defined by
$l=v_{\rm F}\tau_{\rm n}=v_{\rm F}/2\Gamma_{\rm n}$.
   From Eqs.\ (\ref{eq:IMP2}) and (\ref{eq:imp-self2}),
we obtain the self-consistency equations for ${\hat \Sigma}$ as
\begin{eqnarray}
\Sigma_{\rm d}(i\omega_n, {\bf r})
   =
-i \Gamma_{\rm n}
\Bigl\langle g(i\omega_n, {\bf r},{\bar {\bf k}}) \Bigr\rangle,
\label{eq:sigma1}
\end{eqnarray}
\begin{eqnarray}
\Sigma_{12}(i\omega_n, {\bf r})
   =
   \Gamma_{\rm n}
\Bigl\langle f(i\omega_n, {\bf r},{\bar {\bf k}}) \Bigr\rangle,
\label{eq:sigma2}
\end{eqnarray}
\begin{eqnarray}
\Sigma_{21}(i\omega_n, {\bf r})
   =
  - \Gamma_{\rm n}
\Bigl\langle f^{\dagger}(i\omega_n, {\bf r},{\bar {\bf k}}) \Bigr\rangle.
\label{eq:sigma3}
\end{eqnarray}

   The self-consistency equation for $\Delta$,
called gap equation, is given as
\begin{equation}
\Delta({\bf r},{\bar {\bf k}})
=\pi T g F({\bar {\bf k}})
\sum_{-\omega_{\rm c} < \omega_n < \omega_{\rm c}}
\Bigl\langle F^{*}({\bar {\bf k}}') f(i\omega_n, {\bf r},{\bar {\bf k}}')
\Bigr\rangle,
\label{eq:gap}
\end{equation}
where the cutoff energy is $\omega_{\rm c}$,
the pairing interaction is defined as
$g F({\bar {\bf k}}) F^{*}({\bar {\bf k}}')$ with  
the coupling constant $g$ given by
\begin{equation}
\frac{1}{g}
=
\ln\Bigl(\frac{T}{T_{{\rm c} 0}} \Bigr)
+ \sum_{0 \le n < (\omega_{\rm c}/\pi T -1)/2}   \frac{2}{2n+1}.
\label{eq:coupling}
\end{equation}
   We define $T_{{\rm c} 0}$ as the superconducting critical temperature
in the absence of impurities.
   Finally, the system of equations to be solved consists of
the Eilenberger equation (\ref{eq:eilen2}) and
the self-consistency equations for
the impurity self-energies
$\bigl[$Eqs.\ (\ref{eq:sigma1})--(\ref{eq:sigma3})$\bigr]$
and 
for the superconducting order parameter
$\bigl[$Eq.\ (\ref{eq:gap})$\bigr]$.

%
\section{$S$-WAVE AND CHIRAL $P$-WAVE VORTEX SYSTEMS}

   In this study,
the system is assumed to be an isotropic two-dimensional conduction layer
perpendicular to the vorticity along the $z$ axis.
   In the circular coordinate system we use
${\bf r}=(r\cos\phi,r\sin\phi)$ and
${\bar{\bf k}}
=(\cos\theta, \sin\theta)$.
   We assume a circular Fermi surface and
${\bf v}_{\rm F}({\bar{\bf k}})
=v_{\rm F}{\bar{\bf k}}
=(v_{\rm F}\cos\theta,v_{\rm F}\sin\theta)$
in the Eilenberger equations (\ref{eq:eilen}) and (\ref{eq:eilen2}).
   The average over the Fermi surface
reads:
$\langle \cdots \rangle = \int_0^{2\pi} \cdots d \theta/2\pi$.

\subsection{Pair Potential}
   The single vortex is situated at the origin ${\bf r}=0$.
   In the case of the $s$-wave pairing,
the pair potential (i.e., the superconducting order parameter)
around the vortex
is expressed as
\begin{equation}
\Delta_{\rm s}({\bf r})=\Delta_{\rm s}(r) e^{i\phi},
\label{eq:op-s}
\end{equation}
where we can take
$\Delta_{\rm s}(r)$ in the right-hand side
to be real
because of axial symmetry of the system.
   The $s$-wave pairing means
$F({\bar {\bf k}})=1$ and the gap equation (\ref{eq:gap}) reads:
\begin{equation}
\Delta_{\rm s}({\bf r})
=\pi T g_{\rm s}
\sum_{-\omega_{\rm c} < \omega_n < \omega_{\rm c}}
 \int_0^{2\pi} \frac{d \theta}{2\pi}
 f(i\omega_n, {\bf r}, \theta),
\label{eq:gap-s}
\end{equation}
where $g_{\rm s}$ follows Eq.\ (\ref{eq:coupling}).

   On the other hand, 
in the case of the chiral $p$-wave pairing,
$F({\bar {\bf k}})={\bar k}_x \pm i{\bar k}_y = \exp(\pm i\theta)$.
   The pair potential around the vortex
has two possible forms
depending on whether
the chirality and vorticity are parallel or
antiparallel each other.\cite{kato01,matsumoto01,Heeb99}
   Thus, there are two kinds of vortex.
   One form is
\begin{eqnarray}
\Delta^{({\rm n})}({\bf r},\theta)
   &=&                                            \nonumber
\Delta_{+}^{({\rm n})}({\bf r}) e^{+i\theta}
+ \Delta_{-}^{({\rm n})}({\bf r}) e^{-i\theta} \\
   &=&
\Delta_{+}^{({\rm n})}(r) e^{i(\theta-\phi)}
+ \Delta_{-}^{({\rm n})}(r) e^{i(-\theta+\phi)},
\label{eq:op-pm}
\end{eqnarray}
where the chirality (related to the phase ``$+\theta$")
and vorticity (``$-\phi$") are antiparallel
(``negative vortex").
   The other form is
\begin{eqnarray}
\Delta^{({\rm p})}({\bf r},\theta)
   &=&                                            \nonumber
\Delta_{+}^{({\rm p})}({\bf r}) e^{+i\theta}
+ \Delta_{-}^{({\rm p})}({\bf r}) e^{-i\theta} \\
   &=&
\Delta_{+}^{({\rm p})}(r) e^{i(\theta+\phi)}
+ \Delta_{-}^{({\rm p})}(r) e^{i(-\theta+3\phi)},
\label{eq:op-pp}
\end{eqnarray}
where the chirality (``$+\theta$") and vorticity (``$+\phi$") are parallel
(``positive vortex").
   We have assumed that $\Delta_{+}^{({\rm n,p})}$ is the dominant component
in both Eqs.\ (\ref{eq:op-pm}) and (\ref{eq:op-pp}),
and the other component $\Delta_{-}^{({\rm n,p})}$
is the minor one induced with smaller amplitudes inside the vortex core.
The axial symmetry allows us to take
$\Delta_{\pm}^{({\rm n,p})}(r)$
to be real
in each second line of Eqs.\ (\ref{eq:op-pm}) and (\ref{eq:op-pp}).
   Far away from the vortex,
the dominant component
$\Delta_{+}^{({\rm n,p})}(r \rightarrow \infty)$
is finite
and
the induced minor one
$\Delta_{-}^{({\rm n,p})}(r \rightarrow \infty) \rightarrow 0$,
namely
\begin{eqnarray}
\Delta^{({\rm n})}(r \rightarrow \infty, \phi;\theta)=
\Delta_{+}^{({\rm n})}(r \rightarrow \infty) e^{i(\theta-\phi)},
\label{eq:op-pm2}
\end{eqnarray}
\begin{eqnarray}
\Delta^{({\rm p})}(r \rightarrow \infty, \phi;\theta)=
\Delta_{+}^{({\rm p})}(r \rightarrow \infty) e^{i(\theta+\phi)}.
\label{eq:op-pp2}
\end{eqnarray}
   In the clean limit,
$\Delta_{+}^{({\rm n,p})}(r \rightarrow \infty)
=\Delta_{\rm BCS}(T)$
$\bigl[\Delta_{\rm BCS}(T)$ is the BCS gap amplitude$\bigr]$.
   The gap equation (\ref{eq:gap}) reads now:
\begin{equation}
\Delta_{\pm}^{({\rm n,p})}({\bf r})
=\pi T g_{\rm p}
\sum_{-\omega_{\rm c} < \omega_n < \omega_{\rm c}}
 \int_0^{2\pi} \frac{d \theta}{2\pi}
 e^{\mp i\theta}
 f(i\omega_n, {\bf r}, \theta),
\label{eq:gap-p}
\end{equation}
where $g_{\rm p}$ follows Eq.\ (\ref{eq:coupling}).

\subsection{Axial Symmetry and Boundary Condition}
 The numerical calculation of
the self-consistent pair potential $\Delta$
and impurity self energy ${\hat \Sigma}$
requires to restrict ourselves to
a finite spatial region,
which we choose axial symmetric with
a cutoff radius $ r_{\rm c} $.
   Therefore, it is necessary
to fix
the values of $\Delta$ and ${\hat \Sigma}$
for $r > r_{\rm c}$ outside the boundary $r=r_{\rm c}$
when solving the Eilenberger equation (\ref{eq:eilen2}).

   We set,
for the pair potential (\ref{eq:op-s}) of the $s$-wave vortex,
\begin{equation}
\Delta_{\rm s}(r>r_{\rm c},\phi)=\Delta_{\rm s}(r=r_{\rm c}) e^{i\phi}.
\label{eq:boundary-s}
\end{equation}
   For the pair potential (\ref{eq:op-pm})
of the chiral negative $p$-wave vortex, we set
\begin{eqnarray}
\Delta^{({\rm n})}(r>r_{\rm c},\phi;\theta)
   =
\Delta_{+}^{({\rm n})}(r=r_{\rm c}) e^{i(\theta-\phi)},
\label{eq:boundary-pm}
\end{eqnarray}
and for the pair potential (\ref{eq:op-pp})
of the chiral positive $p$-wave vortex
\begin{eqnarray}
\Delta^{({\rm p})}(r>r_{\rm c},\phi;\theta)
   =
\Delta_{+}^{({\rm p})}(r=r_{\rm c}) e^{i(\theta+\phi)},
\label{eq:boundary-pp}
\end{eqnarray}
while $\Delta_{-}^{({\rm n,p})}(r>r_{\rm c})=0$.

   Next we consider
the symmetry property and boundary condition
of the impurity self energy ${\hat \Sigma}$.
   In the Eilenberger equation (\ref{eq:eilen2}),
${\hat \Sigma}$ appears in 
Eqs.\ (\ref{eq:self-w}) and (\ref{eq:self-delta}) in the form of
 $i\omega_n - \Sigma_{\rm d}$,
$\Delta + \Sigma_{12}$,
and
$\Delta^{*} - \Sigma_{21}$.
   Owing to the axial symmetry of our model,
an axial rotation of the system
($\phi \rightarrow \phi + \alpha$ and $\theta \rightarrow \theta + \alpha$)
leads to a transformation of the impurity self energies
$\Sigma_{\rm d}$, $\Sigma_{12}$, and $\Sigma_{21}$
in
the same manner as $i\omega_n$, $\Delta$, and $\Delta^{*}$, respectively.
   The Matsubara frequency $i\omega_n$ is invariant under the axial rotation,
and therefore $\Sigma_{\rm d}(i\omega_n, r,\phi)$
has no azimuthal $\phi$ dependence. Hence,
we set
\begin{eqnarray}
\Sigma_{\rm d}(i\omega_n, r,\phi)
   =
\Sigma_{\rm d}(i\omega_n, r),
\label{eq:Sig-d}
\end{eqnarray}
\begin{eqnarray}
\Sigma_{\rm d}(i\omega_n,r>r_{\rm c})
   =
\Sigma_{\rm d}(i\omega_n,r=r_{\rm c}),
\label{eq:boundary-Sig-d}
\end{eqnarray}
both in the case of the $s$-wave vortex and the chiral $p$-wave vortex.

For the $s$-wave vortex, 
the pair potential
$\Delta_{\rm s}({\bf r})$ in Eq.\ (\ref{eq:op-s})
transforms under an axial rotation as
$\Delta_{\rm s}(r,\phi+\alpha)
=\Delta_{\rm s}(r) \exp\bigl[i(\phi+\alpha)\bigr]
=\Delta_{\rm s}(r,\phi) \exp(i\alpha)$,
and
$\Delta_{\rm s}^{*}(r,\phi+\alpha)
=\Delta_{\rm s}^{*}(r,\phi) \exp(-i\alpha)$.
   This rotation means that
$\Delta_{\rm s}+\Sigma_{\rm 12}
\rightarrow (\Delta_{\rm s}+\Sigma_{\rm 12}) \exp(i\alpha)$
and
$\Delta_{\rm s}^{*}-\Sigma_{\rm 21}
\rightarrow (\Delta_{\rm s}^{*}-\Sigma_{\rm 21}) \exp(-i\alpha)$. 
   Thus, the off-diagonal impurity self energies
$\Sigma_{\rm 12,21}(i\omega_n, r,\phi)$
have to possess an azimuthal $\phi$ dependence as
\begin{equation}
\Sigma_{\rm 12}(i\omega_n, r,\phi)
=\Sigma_{\rm 12}(i\omega_n, r)\exp(i\phi),
\label{eq:Sig-12-s}
\end{equation}
\begin{equation}
\Sigma_{\rm 21}(i\omega_n, r,\phi)
=\Sigma_{\rm 21}(i\omega_n, r)\exp(-i\phi).
\label{eq:Sig-21-s}
\end{equation}
   We set, for the $s$-wave vortex, the boundary condition as
\begin{equation}
\Sigma_{\rm 12}(i\omega_n, r>r_{\rm c})
=
\Sigma_{\rm 12}(i\omega_n, r=r_{\rm c}),
\label{eq:boundary-Sig-12-s}
\end{equation}
\begin{equation}
\Sigma_{\rm 21}(i\omega_n, r>r_{\rm c})
=
\Sigma_{\rm 21}(i\omega_n, r=r_{\rm c}),
\label{eq:boundary-Sig-21-s}
\end{equation}
because far away from the vortex core
the anomalous Green functions averaged over the Fermi surface,
which appear in Eqs.\ (\ref{eq:sigma2}) and (\ref{eq:sigma3}),
are generally nonzero
and their amplitudes are spatially uniform,
owing to the $s$-wave pairing symmetry.\cite{haya03-1}

   For the chiral negative $p$-wave vortex,
the pair potential
$\Delta^{({\rm n})}$
has the following symmetry
properties. For an axial rotation,
the pair potential
$\Delta^{({\rm n})}({\bf r},\theta)$ in Eq.\ (\ref{eq:op-pm}) transforms
as
$\Delta^{({\rm n})}(r,\phi+\alpha;\theta+\alpha)
=\Delta^{({\rm n})}(r,\phi;\theta)$
and
$\Delta^{({\rm n}) *}(r,\phi+\alpha;\theta+\alpha)
=\Delta^{({\rm n}) *}(r,\phi;\theta)$,
namely invariant.
   Therefore, we find that 
$\Delta^{({\rm n})}+\Sigma_{\rm 12}
\rightarrow \Delta^{({\rm n})}+\Sigma_{\rm 12}$
and
$\Delta^{({\rm n}) *}-\Sigma_{\rm 21}
\rightarrow \Delta^{({\rm n}) *}-\Sigma_{\rm 21}$
and 
the off-diagonal impurity self energies
$\Sigma_{\rm 12,21}(i\omega_n, r,\phi)$ are not
$\phi$-dependent:
\begin{equation}
\Sigma_{\rm 12}(i\omega_n, r,\phi)
=\Sigma_{\rm 12}(i\omega_n, r),
\label{eq:Sig-12-pm}
\end{equation}
\begin{equation}
\Sigma_{\rm 21}(i\omega_n, r,\phi)
=\Sigma_{\rm 21}(i\omega_n, r).
\label{eq:Sig-21-pm}
\end{equation}
   We set, for the chiral negative $p$-wave vortex, the boundary condition as
\begin{equation}
\Sigma_{\rm 12}(i\omega_n, r>r_{\rm c})
=
0,
\label{eq:boundary-Sig-12-pm}
\end{equation}
\begin{equation}
\Sigma_{\rm 21}(i\omega_n, r>r_{\rm c})
=
0,
\label{eq:boundary-Sig-21-pm}
\end{equation}
because far away from the vortex core
the anomalous Green functions averaged over the Fermi surface
are zero, owing to the $p$-wave pairing symmetry.\cite{haya03-1}

   On the other hand, the chiral positive $p$-wave vortex
behaves differently under
axial rotation. 
The pair potential
$\Delta^{({\rm p})}({\bf r},\theta)$ in Eq.\ (\ref{eq:op-pp}) transforms
as
$\Delta^{({\rm p})}(r,\phi+\alpha;\theta+\alpha)
=\Delta^{({\rm p})}(r,\phi;\theta)\exp(2i\alpha)$
and
$\Delta^{({\rm p}) *}(r,\phi+\alpha;\theta+\alpha)
=\Delta^{({\rm p}) *}(r,\phi;\theta)\exp(-2i\alpha)$,
such that 
$\Delta^{({\rm p})}+\Sigma_{\rm 12}
\rightarrow (\Delta^{({\rm p})}+\Sigma_{\rm 12})\exp(2i\alpha)$
and
$\Delta^{({\rm p}) *}-\Sigma_{\rm 21}
\rightarrow (\Delta^{({\rm p}) *}-\Sigma_{\rm 21})\exp(-2i\alpha)$.
Here, the off-diagonal impurity self energies
$\Sigma_{\rm 12,21}(i\omega_n, r,\phi)$
depends on $\phi$ like
\begin{equation}
\Sigma_{\rm 12}(i\omega_n, r,\phi)
=\Sigma_{\rm 12}(i\omega_n, r) \exp(2i\phi),
\label{eq:Sig-12-pp}
\end{equation}
\begin{equation}
\Sigma_{\rm 21}(i\omega_n, r,\phi)
=\Sigma_{\rm 21}(i\omega_n, r) \exp(-2i\phi).
\label{eq:Sig-21-pp}
\end{equation}
   We set, for the chiral positive $p$-wave vortex, the boundary condition as
\begin{equation}
\Sigma_{\rm 12}(i\omega_n, r>r_{\rm c})
=
0,
\label{eq:boundary-Sig-12-pp}
\end{equation}
\begin{equation}
\Sigma_{\rm 21}(i\omega_n, r>r_{\rm c})
=
0,
\label{eq:boundary-Sig-21-pp}
\end{equation}
as in the case of the negative vortex.

\section{VORTEX CORE RADIUS $\xi_{1}(T)$}
   The vortex core radius $\xi_1(T)$ defined in Eq.\ (\ref{eq:KP})
is obtained from the spatial profile of the pair potential
calculated self-consistently.
   We solve the Eilenberger equation (\ref{eq:eilen2})
by a method of the Riccati parametrization,\cite{schopohl2,schopohl}
and then iterate the calculation, until self-consistency is reached,
with the self-consistency equations
of the impurity self energies
$\bigl[$Eqs.\ (\ref{eq:sigma1})--(\ref{eq:sigma3})$\bigr]$
and that of the pair potential
for the $s$-wave vortex
$\bigl[$Eq.\ (\ref{eq:gap-s})$\bigr]$
or
for the chiral $p$-wave vortices
$\bigl[$Eq.\ (\ref{eq:gap-p})$\bigr]$.    
We use an acceleration method for iterative calculations
to obtain sufficient accuracy.\cite{eschrig1}
   When solving the Eilenberger equation (\ref{eq:eilen2}),
we use the boundary conditions for the pair potential
and the impurity self energies described in Sec.\ 3.2.

   We show now the results of $\xi_1(T)$ for the three cases 
   introduced above.
   We set the cutoff energy $\omega_{\rm c}=10 \Delta_0$
and the cutoff length $r_{\rm c}=10 \xi_0$.
   Here, $\Delta_0$ is the BCS gap amplitude at zero temperature
in the absence of impurities,
and $\xi_0=v_{\rm F}/\Delta_0$.
   The lowest value of the temperature at which $\xi_1$ is computed
is $T=0.02T_{{\rm c}0}$.

   We have checked an influence of the finite system size,
comparing results of $\xi_1$
obtained for
two cutoff lengths
$r_{\rm c}=10 \xi_0$ and $20 \xi_0$
in the case of the clean limit.
   As a result, it was found that
those results well coincide each other
below $T \sim 0.7 T_{{\rm c}0}$
(the deviations in $\xi_1$ are less than
$0.003 \xi_0$ and negligibly small),
while slight deviations appear
above $T \sim 0.8 T_{{\rm c}0}$,
but are less than
$0.014 \xi_0$
($T \le 0.9 T_{{\rm c}0}$)
and almost invisible for plots
with the same plot scale as in the following figures.

\subsection{$S$-Wave Vortex}
\begin{figure}
\begin{center}
\includegraphics[scale=0.79]{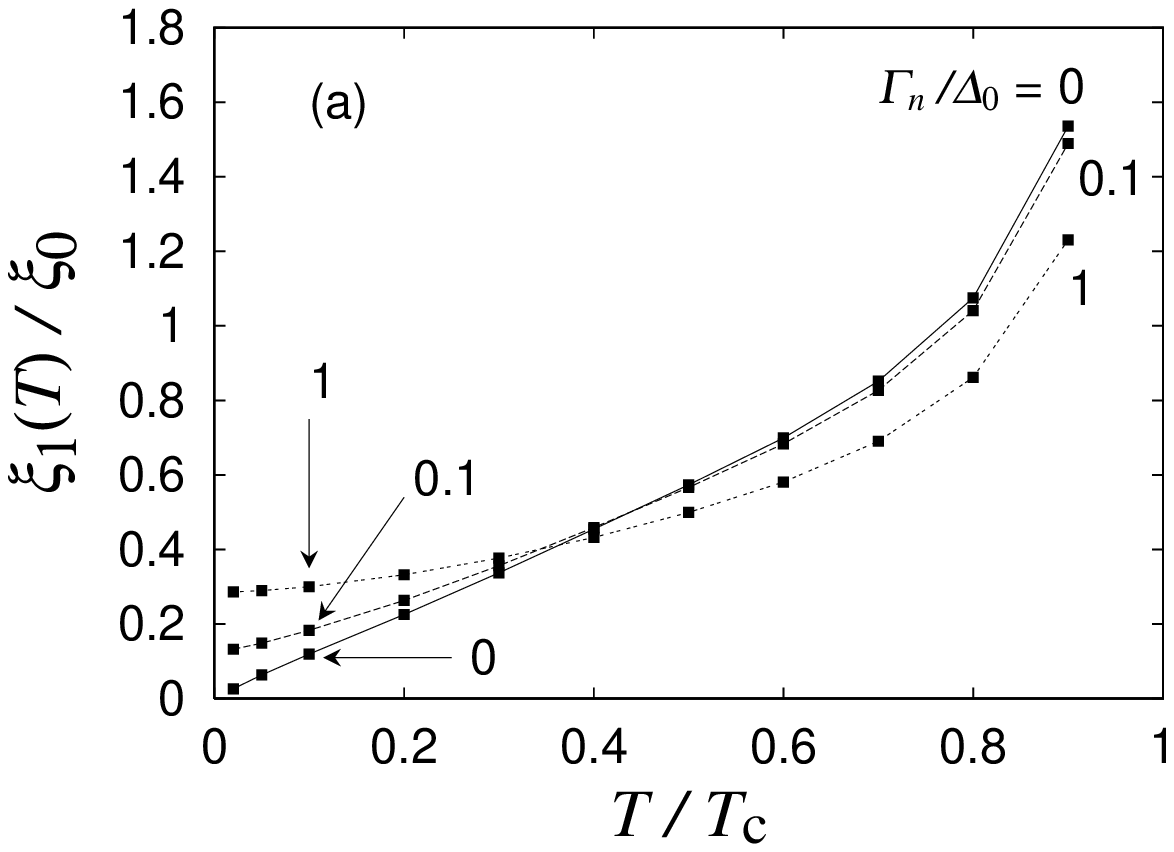}
\includegraphics[scale=0.79]{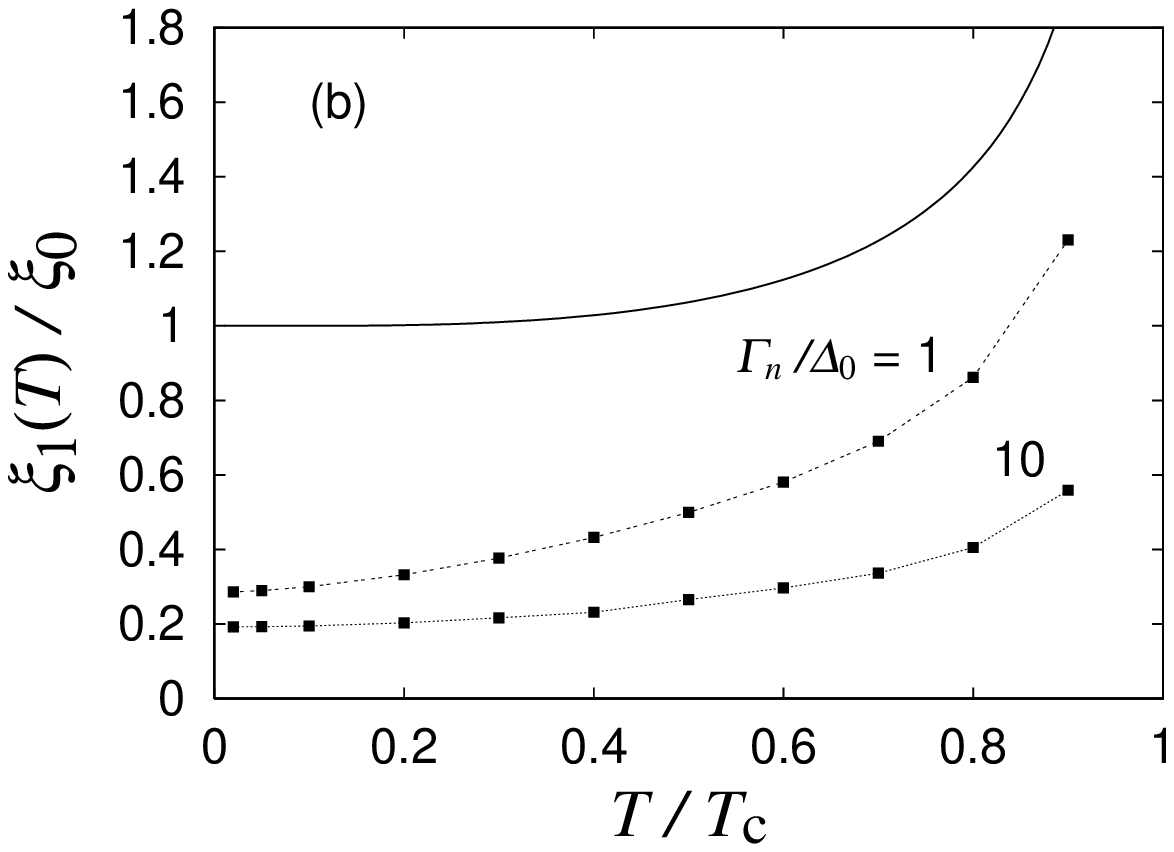}
\end{center}
\caption{
   The vortex core radius $\xi_1(T)$ (points)
in the case of the $s$-wave vortex (\ref{eq:op-s})
as a function of the temperature $T$
for several values of the impurity scattering rate $\Gamma_{\rm n}$.
   Lines are guides for the eye, except for the solid line in (b).
   In the plots, $\xi_1$ and $T$ are normalized by
$\xi_0$ and $T_{\rm c}$, respectively.
   Here, $T_{\rm c}$ is the superconducting critical temperature,
$\xi_0$ is defined as $\xi_0=v_{\rm F}/\Delta_0$,
and $\Delta_0$ is the BCS gap amplitude at zero temperature
in the absence of impurities.
(a) $\Gamma_{\rm n} = 0$, 0.1$\Delta_0$, and $\Delta_0$.
(b) $\Gamma_{\rm n} = \Delta_0$ and $10\Delta_0$.
   The solid line in (b)
is a plot of
the function,
$1/\tanh \bigl(1.74\sqrt{(T_{\rm c}/T)-1}\bigr)
\approx \Delta_0/\Delta_{\rm BCS}(T)
=v_{\rm F}/\Delta_{\rm BCS}(T)\xi_0$,
which reproduces approximately the temperature dependence
of the clean-limit BCS coherence length.
}
\label{fig:1}
\end{figure}
   In Fig.\ 1,
we show the vortex core radius $\xi_1(T)$ for the $s$-wave vortex
as a function of the temperature $T$
for several values of the impurity scattering rate $\Gamma_{\rm n}$.
 The critical temperature $T_{\rm c}$
remains unaffected by the nonmagnetic impurities,
namely $T_{\rm c} = T_{{\rm c}0}$ for any values of $\Gamma_{\rm n}$.

   At high temperatures,
we see in Fig.\ 1 that
the vortex core radius $\xi_1$ decreases
with the increase of
the impurity scattering rate $\Gamma_{\rm n}$.
   This is because
the coherence length $\xi$,
over which the pair potential significantly changes,
shrinks 
with the decrease of
the quasiparticle mean free path
($\propto 1/\Gamma_{\rm n}$).\cite{tinkham}
   The coherence length $\xi$
is a distance
at which
the pair potential
is restored
far away from the vortex center.
   The vortex core radius $\xi_1$ defined
in the vicinity of the vortex center
is dominated
by this $\Gamma_{\rm n}$ dependence
of the coherence length $\xi$ in this temperature regime.

   Pronounced $\Gamma_{\rm n}$ dependence
appears also at low temperatures.
   For $\Gamma_{\rm n}=0$ (the clean limit),
the vortex core radius $\xi_1$
decreases linearly in $T$,
as expected for the KP effect.
   In Fig.\ 1(a), we also show $\xi_1(T)$
for finite values of the impurity scattering rate,
$\Gamma_{\rm n}=0.1 \Delta_0$ and $\Delta_0$.
    At low temperatures,
the vortex core radius $\xi_1$
{\it increases}
with the increase of $\Gamma_{\rm n}$, in contrast to
the high-temperature behavior.
   This increase of $\xi_1$
indicates the saturation feature of the KP effect
due to impurities.\cite{KP}
   The low-temperature vortex core radius $\xi_1$ 
expands by introducing impurity scattering but still
remains much smaller than $ \xi $. 
   For relatively small $\Gamma_{\rm n}$ $(< \Delta_0)$,
the decrease of the coherence length $\xi$
mentioned above
has little influence
on this expansion of the vortex core radius $\xi_1$ ($\ll \xi$).
   For larger $\Gamma_{\rm n}$ towards the dirty limit, however,
the decrease of the coherence length $\xi$
begins to influence
the vortex core radius $\xi_1$,
and $\xi_1$ begins to decrease with growing $\Gamma_{\rm n}$
as seen in Fig.\ 1(b).

   In Fig.\ 1(b), the solid line displays
the function,
$1/\tanh \bigl(1.74\sqrt{(T_{\rm c}/T)-1}\bigr)$
$\approx \Delta_0/\Delta_{\rm BCS}(T)
=v_{\rm F}/\Delta_{\rm BCS}(T)\xi_0$.
   In the dirty case
($\Gamma_{\rm n}=10\Delta_0$,
i.e., the mean free path
$l=v_{\rm F}/2\Gamma_{\rm n}=0.05\xi_0$),
$\xi_1(T)$
behaves like the clean-limit BCS coherence length
$\sim v_{\rm F}/\Delta_{\rm BCS}(T)$
below $T \sim 0.6 T_{\rm c}$,
and
is almost constant at low temperatures.
   The increase of the vortex core radius $\xi_1(T)$
with increasing $T$
at high temperatures for the dirty case $\Gamma_{\rm n}=10 \Delta_0$,
is more gradual than the temperature dependence of
the clean-limit BCS coherence length
$\bigl($the solid line in Fig.\ 1(b)$\bigr)$.
   This behavior is qualitatively consistent with a dirty limit result
reported by Volodin {\it et al.}\cite{golubov97}
Note that
the overall temperature dependence of 
$\xi_1$ in the dirty case
is quantitatively different
from that of Ref.\ \onlinecite{golubov97} ($\xi_{\rm eff}$),
which is probably caused by the difference of  the definitions of the vortex core radius.
   As displayed in Fig.\ 1(a) for the
moderately clean case
($\Gamma_{\rm n}=0.1\Delta_0$,
i.e., $l=5\xi_0$)
and even in the relatively dirty case
($\Gamma_{\rm n}=\Delta_0$,
i.e., $l=0.5\xi_0$),
the vortex core radius $\xi_1(T)$
shrinks
approximately linearly in $T$
with moderate curvature
below $T \sim 0.6 T_{\rm c}$
and saturates towards a finite value in zero-temperature limit.
   This gradual saturation
due to impurities
is
in contrast to
a sudden truncation of the KP effect
which happens
below a certain temperature related to
the discrete energy levels of
the low-lying vortex bound states.\cite{haya98,m-kato}

\subsection{Chiral $P$-Wave Vortices}
\begin{figure}
\begin{center}
\includegraphics[scale=0.79]{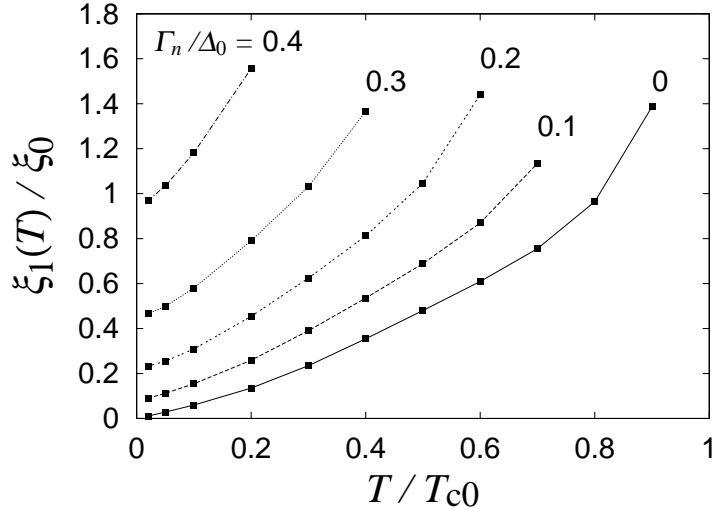}
\end{center}
\caption{
   The vortex core radius $\xi_1(T)$ (points)
in the case of
the chiral {\it positive} $p$-wave vortex (\ref{eq:op-pp})
as a function of the temperature $T$
for several values of the impurity scattering rate $\Gamma_{\rm n}$.
   Lines are guides for the eye.
   In the plot, $\xi_1$ and $T$ are normalized by
$\xi_0$ and $T_{{\rm c}0}$, respectively.
   Here, $T_{{\rm c}0}$ is the superconducting critical temperature
in the absence of impurities,
$\xi_0$ is defined as $\xi_0=v_{\rm F}/\Delta_0$,
and $\Delta_0$ is the BCS gap amplitude at zero temperature
in the absence of impurities.
}
\label{fig:2}
\end{figure}
   In the chiral $p$-wave superconductors
and generally in unconventional superconductors,
the superconducting critical temperature
decreases in the presence of impurities.
   The units of the temperature $T$ in
Figs.\ 2 and 3 is $T_{{\rm c}0}$
(the superconducting critical temperature in the absence of impurities).
   We obtain $\xi_1$
from the dominant components $\Delta_{+}^{({\rm n,p})}(r)$
in Eqs.\ (\ref{eq:op-pm}) and (\ref{eq:op-pp}).

\begin{figure}
\begin{center}
\includegraphics[scale=0.79]{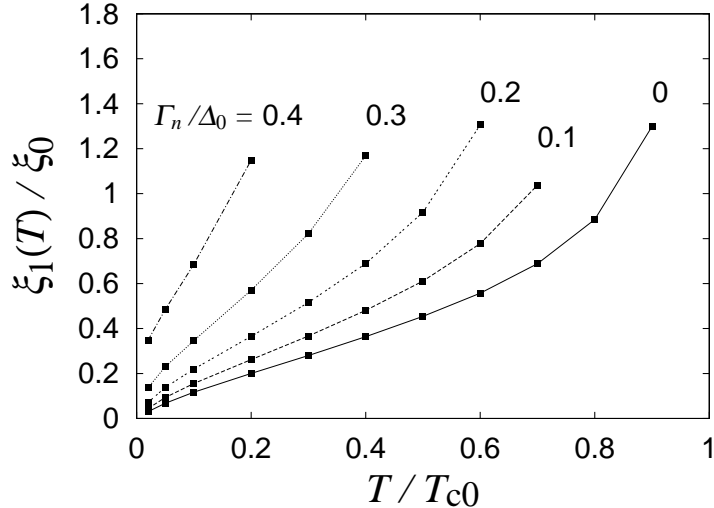}
\end{center}
\caption{
   The vortex core radius $\xi_1(T)$ (points)
in the case of
the chiral {\it negative} $p$-wave vortex (\ref{eq:op-pm})
as a function of the temperature $T$
for several values of the impurity scattering rate $\Gamma_{\rm n}$.
   Lines are guides for the eye.
   In the plot, $\xi_1$ and $T$ are normalized by
$\xi_0$ and $T_{{\rm c}0}$, respectively.
   Here, $T_{{\rm c}0}$ is the superconducting critical temperature
in the absence of impurities,
$\xi_0$ is defined as $\xi_0=v_{\rm F}/\Delta_0$,
and $\Delta_0$ is the BCS gap amplitude at zero temperature
in the absence of impurities.
}
\label{fig:3}
\end{figure}
   We show in Fig.\ 2
the vortex core radius $\xi_1(T)$
in the case of the positive vortex
$\Delta^{(\rm p)}$
$\bigl[$Eq.\ (\ref{eq:op-pp})$\bigr]$.
   At low temperatures, $\xi_1(T)$ shrinks
approximately linearly in $T$
with moderate curvature
and saturates towards a finite value in zero-temperature limit,
as in the case of the $s$-wave vortex.
   The vortex core radius $\xi_1$ expands
with the increase of $\Gamma_{\rm n}$,
owing to the increase of
the coherence length $\xi$ with the suppression of
the pair potential far away from the vortex core.

   Figure 3 displays $\xi_1(T)$ for the negative vortex
$\Delta^{(\rm n)}$
$\bigl[$Eq.\ (\ref{eq:op-pm})$\bigr]$.
   In contrast to the cases of the $s$-wave vortex
and the positive vortex,
the vortex core radius
$\xi_1(T)$ strongly decreases
even in the presence of impurities,
and shrinks toward almost zero
for the moderately clean cases
below $\Gamma_{\rm n} \sim 0.2\Delta_0$.

   We can understand,
in the following way,
the reason why
the vortex core shrinkage is robust against impurities. 
   In the Eilenberger equations (\ref{eq:eilen}) and (\ref{eq:eilen2}),
the term which includes the impurity self energy is written as
\begin{equation}
\bigl[ {\hat \sigma},{\hat g} \bigr]
=
\bigl[ {\hat \Sigma},{\hat g} \bigr]
=
\frac{n_{\rm i} N_0 v^2}{D}\bigl[ \langle{\hat g}\rangle,{\hat g} \bigr],
\label{eq:imp-term}
\end{equation}
where we referred to Eqs.\ (\ref{eq:IMP3}) and (\ref{eq:imp-self}).
   Therefore, if the Fermi-surface-averaged Green function
$\langle{\hat g}\rangle$ is equivalent to
the original Green function ${\hat g}$
$\bigl($namely, if $\langle{\hat g}\rangle = {\hat g}$$\bigr)$,
the above term (\ref{eq:imp-term})
becomes zero and the impurity does not play a role
in the Eilenberger equation.
   Such a special situation occurs in the negative vortex (\ref{eq:op-pm})
of the chiral $p$-wave phase,
and not in the positive vortex (\ref{eq:op-pp})
or other usual vortices.\cite{haya02-1,haya03-1}
   It corresponds to a local restoration of the Anderson's theorem
inside the vortex core.\cite{volovik99,kato00,kato02,haya02-1,haya03-1,kato03,matsumoto00}
   This negative vortex (\ref{eq:op-pm})
 is more favorable energetically
than the positive vortex (\ref{eq:op-pp})
at least in the clean limit,\cite{Heeb99}
and therefore the negative vortex is likely to exist in
the chiral $p$-wave superconductors.

\section{SUMMARY}
   We investigated the temperature dependence of
the vortex core radius $\xi_1(T)$ defined in Eq.\ (\ref{eq:KP}),
incorporating
the effect of nonmagnetic impurities in the Born limit.
   The isolated single vortex in the isotropic two-dimensional system
was considered
for the $s$-wave pairing symmetry
and the chiral $p$-wave pairing symmetry. 

   In the case of the $s$-wave vortex (\ref{eq:op-s}),
as seen in Fig.\ 1,
{\it at low temperatures}
the vortex core radius $\xi_1$
{\it increases}
with the increase of the impurity scattering rate $\Gamma_{\rm n}$
up to $\Gamma_{\rm n} \sim \Delta_0$
owing to the saturation of the KP effect due to impurities.
   In contrast to it,
{\it at high temperatures} the vortex core radius $\xi_1$
{\it decreases}
with the increase of $\Gamma_{\rm n}$
owing to the decrease of the coherence length $\xi$ due to impurities.
   In the case of
the $s$-wave vortex (\ref{eq:op-s}) of the moderately clean state
($\Gamma_{\rm n} = 0.1\Delta_0$,
i.e., $l = 5\xi_0$)
and of the relatively dirty state
($\Gamma_{\rm n}=\Delta_0$,
i.e., $l=0.5\xi_0$),
as seen in Figs.\ 1(a),
at low temperatures
the vortex core radius $\xi_1(T)$
{\it shrinks approximately linearly in $T$
with moderate curvature}
and saturates towards a finite value in zero-temperature limit.
   This gradual saturation
due to impurities
is in contrast to
a sudden truncation of the KP effect due to the discreteness in
the energy spectrum of the low-lying vortex bound states.\cite{haya98,m-kato}
   In the case of the chiral $p$-wave pairing system,
as seen in Fig.\ 2,
for the positive vortex (\ref{eq:op-pp})
the shrinkage of
the vortex core radius $\xi_1(T)$ saturates towards a finite value
in zero-temperature limit
by impurity scattering,
analogous to the $s$-wave vortex.
   For the negative vortex (\ref{eq:op-pm}), however,
the local restoration of
the Anderson's theorem
inside the vortex core\cite{volovik99,kato00,kato02,haya02-1,haya03-1,kato03,matsumoto00}
yields a KP effect little affected by impurity scattering
and,
as seen in Fig.\ 3,
the vortex core radius $\xi_1(T)$
{\it strongly shrinks
linearly in $T$
at low temperatures
even in the presence of impurities}.

   It would naturally be highly desirable to establish
the KP effect experimentally
beyond the present level.
   Our analysis shows that impurity scattering which is
harmful for the KP effect in conventional superconductors,
can under certain 
condition be harmless in a chiral $p$-wave superconductor,
making this kind of
system a good candidate for experimental tests.
   On the other hand,
we expect that
rather weak shrinkings of vortex core observed
in NbSe$_2$ (Refs.\ \onlinecite{sonier04-1,sonier00,sonier97,miller})
and CeRu$_2$ (Ref.\ \onlinecite{kadono01})
may be explained partly
in terms of
the impurity effect,
i.e.,
the vortex core shrinkage
approximately linear in $T$
with moderate curvature
and its saturation towards a finite value
in zero-temperature limit $\bigl($Fig.\ 1(a)$\bigr)$.
   Rather large extrapolated values of
the vortex core radius at zero temperature
observed experimentally
might be partly attributed to effects of multiple gaps.
   Finally,
we mention a multi-gap effect on the KP effect.
   Our preliminary results for $r_0 (T)$
in a two-gap model
show that
a contribution from
the Fermi surface with a smaller gap to the total supercurrent density
$j(r)$ around a vortex
makes the position $r_0$, at which $|j(r)|$ has its maximum value,
shift outward away from the vortex center $r=0$,
leading to a finite $r_0$ at zero temperature
in spite of the clean limit.\cite{haya}
   Our detailed results for the KP effect in a two-gap superconductor
will be reported elsewhere.
   Sr$_2$RuO$_4$ (Refs.\ \onlinecite{maeno,agterberg,kusunose2})
and NbSe$_2$ (Ref.\ \onlinecite{nbse2})
have multiple bands and may be effectively two-gap superconductors.
   MgB$_2$ is a typical two-gap
superconductor.\cite{choi02}
   Further investigations on the vortex core shrinkage
in terms of the multi-gap effects
are left for future experimental and theoretical studies.

\section*{ACKNOWLEDGMENTS}
   We would like to thank N.\ Schopohl, T.\ Dahm, and S. Graser
for enlightening discussions.
   One of the authors (N.H.) is grateful for the support by
2003 JSPS Postdoctral Fellowships for Research Abroad. 
   We acknowledge gratefully financial support from the Swiss Nationalfonds.


\end{document}